# Interference induced enhancement of magneto-optical Kerr effect in ultrathin magnetic films

Satoshi Sumi[1], Hiroyuki Awano[1] and Masamitsu Hayashi[2,3]*

[1] *Toyota Technological Institute, Nagoya 468-8511, Japan*

[2] *Department of Physics, The University of Tokyo, Bunkyo, Tokyo 113-0033, Japan*

[3] *National Institute for Materials Science, Tsukuba 305-0047, Japan*

We have studied the magneto-optical spectra of ultrathin magnetic films deposited on Si substrates coated with an oxide layer (SiOx). We find that the Kerr rotation angle and the ellipticity of ~1 nm thick CoFeB thin films, almost transparent to visible light, show a strong dependence on the thickness of the SiOx layer. The Kerr signal from the 1 nm CoFeB thin film can be larger than that of ~100 nm thick CoFeB films for a given SiOx thickness and light wavelength. The enhancement of the Kerr signal occurs when optical interference takes place within the SiOx layer. Interestingly, under such resonance condition, the measured Kerr signal is in some cases larger than the estimation despite the good agreement of the measured and calculated reflection amplitude. We infer the discrepancy originates from interface states that are distinct from the bulk characteristics. These results show that optical interference effect can be utilized to study the magneto-optical properties of ultrathin films.

*Email: hayashi@phys.s.u-tokyo.ac.jp

**Introduction**

Thin film heterostructures with ultrathin magnetic layer(s) are attracting great interest owing to the strong interfacial effects found recently[1,2]. The perpendicular magnetic anisotropy at the interface between a magnetic layer and an oxide layer has allowed development of magnetic storage elements for the spin transfer torque-magnetic random access memory (STT-MRAM) applications[3-5]. Layers with strong spin orbit interaction (e.g. 5$d$ transition metals) can generate spin current and/or spin accumulation, which allows current controlled magnetization of the thin magnetic layer via spin orbit torques[6-9]. Chiral magnetic structure emerges in similar heterostructures due to the interface Dzyaloshinskii-Moriya interaction[10,11].

Magneto-optical Kerr effect provides a non-destructive approach to gaining information of the magnetization direction of magnetic materials[12-15]. It has been widely used recently to probe the magnetic state of thin film heterostructures to evaluate spin-orbit torques, chiral magnetism and related effects[16-25]. More recently, the Kerr effect has also been used to reveal the magnetization of two-dimensional (2D) monoatomic van der Waals crystals[26,27]. As the magnetic layer in many of the thin film heterostructures is only a few atomic layers thick and hardly reflects light, it is of interest to study how one can increase the Kerr signal from the heterostructures to allow evaluation of the magnetic states.

Optical interference effect has been commonly used to increase the Kerr signal from magnetic materials[28-34]. Dielectric materials with designed thickness (for a given wavelength) have been placed above or beneath the magnetic material so that multiple reflections occur at the top and bottom interfaces of the dielectric layer. If many reflections take place at the dielectric/magnetic layer interface, the effective Kerr rotation angle can be increased. Insertion of proper dielectric materials can boost the Kerr signal by more than a factor of ten[29,33] and has been incorporated into magneto-optical storage technologies[35]. However, this effect has not been confirmed for ultrathin magnetic layers which are almost transparent to visible light.

Here we study the magneto-optical Kerr spectrum of ultrathin magnetic films deposited on Si substrates coated with silicon oxide (SiOx) to identify the origin of the large Kerr signals found in such heterostructures. Even for films with magnetic layer of ~1 nm thickness, which is almost transparent to visible light, we find that the Kerr rotation angle and its ellipticity can be significantly enhanced by the thickness of SiOx for a given light wavelength. The Kerr signal of the ultrathin magnetic films is larger than that of the corresponding bulk value under certain conditions. The increase in the Kerr rotation angle and the ellipticity can be described with an analytical model that takes into account optical interference effect[13,36-38] within the SiOx layer. These results show that the magnetic state of any ultrathin magnetic films, which hardly reflect light in the visible range, can be evaluated using the Kerr effect by properly choosing the substrate and the light wavelength.

**Results**

Films are deposited using RF magnetron sputtering. The heterostructure used here to study the magneto-optical Kerr effect is: Sub./0.5 Ta/1 CoFeB/2 MgO/1 Ta (thickness in nanometer). We use non-doped Si (100) substrates coated with SiOx (thickness is $t_{SiOx}$). The heterostructures are annealed at ~300 °C ex-situ in vacuum for 1 hour. The light wavelength dependence of the Kerr rotation angle and ellipticity are measured using a homemade Kerr spectroscopy. The refractive index of the substrate and the layers constituting the heterostructure are studied using an ellipsometry. We measure the refractive index of ~100 nm thick single layer films which we assume represent the dielectric properties of the corresponding layer in the heterostructure. See the Methods section for the details of the sample preparation and experimental setup.

Figure 1(a) shows the magnetization hysteresis loops obtained by sweeping the magnetic field directed along the film normal while the polar Kerr signal (*A*) is measured. As schematically shown to the right of Fig. 1(a), the light is incident from near film normal. The wavelength of the light is fixed to ~500 nm here. Clear Kerr signal from the magnetization of the CoFeB layer is found except for the film without the $SiO_X$ layer ($t_{SiOx}$=0 nm). The amplitude of the Kerr rotation ($\theta_K$) is defined as the difference in the signal when the magnetization ($m_Z$) is reversed along its easy axis, i.e. $\theta_K \equiv A(m_Z = -1) - A(m_Z = 1)$. $\theta_K$ is plotted as a function of $t_{SiOx}$ in Fig. 1(b): the amplitude varies with the thicknesses of the SiOx layer.

The wavelength dependence of the reflection amplitude is shown in Figs. 2(a-g) for the films with different $t_{SiOx}$. The refection amplitude shows an oscillatory dependence on the wavelength when $t_{SiOx}$ is non-zero. The corresponding Kerr rotation angle ($\theta_K$) and the ellipticity ($\eta_K$) are plotted in Figs. 3(a-g) with black squares and red circles, respectively. The ellipticity is defined in units of angle. $\theta_K$ and $\eta_K$ exhibit symmetric and/or asymmetric peak-like structures when the reflection amplitude takes a minimum. In order to account for the origin of the Kerr spectra's peaks, we calculate $\theta_K$ and $\eta_K$ using the effective refractive index approach[36]. Note that the matrix approach[13] returns the same results.

The refractive index of a double layer system $\beta/\alpha$ ($\alpha$ is sitting on top of $\beta$) can be modeled as a single layer medium with an effective refractive index ($n_{\beta/\alpha}, k_{\beta/\alpha}$), which is defined as[36]:

$$n_{\beta/\alpha} - ik_{\beta/\alpha} = (n_\alpha - ik_\alpha) \frac{1 - r_{\beta\alpha} \exp\left[2i(n_\alpha - ik_\alpha)\frac{2\pi t_\alpha}{\lambda}\right]}{1 + r_{\beta\alpha} \exp\left[2i(n_\alpha - ik_\alpha)\frac{2\pi t_\alpha}{\lambda}\right]} \qquad (1)$$

Here $\lambda$ is the wavelength of the light incident from medium $\alpha$. $t_\alpha$ is the thickness of medium $\alpha$. $n_l$ and $k_l$ ($l= \alpha, \beta$) are, respectively, the real and imaginary components of the refractive index of medium $l$. $r_{\beta\alpha}$ is the reflection amplitude of the light at the $\beta/\alpha$ interface and is defined as:

$$r_{\beta\alpha} = \frac{(n_\alpha - ik_\alpha) - (n_\beta - ik_\beta)}{(n_\alpha - ik_\alpha) + (n_\beta - ik_\beta)} \qquad (2)$$

To calculate the Kerr spectrum, we need values of the refractive index for the relevant layers. Here we assume that our system constitutes from three media, i.e., the Si substrate, the SiOx layer, and the thin film heterostructure. Although the thin film heterostructure includes four different layers, the dominant contribution to the magneto-optical properties arises from the magnetic (CoFeB) layer. The other layers are thin enough, compared to the light wavelength, such that interference effects within each layer can be neglected. In using Eqs. (1) and (2) one must know the refractive index of three media.

The real ($n_{Si}$) and imaginary ($k_{Si}$) parts of the refractive index of the Si substrate without any SiOx layer ($t_{SiOx}$~0 nm) is plotted in Fig. 4(a). A peak is found at low wavelength, a characteristic well known for Si (ref. [39]). For the SiOx layer, we take a value ($n_{SiOx}$=1.5, $k_{SiOx}$=0) widely used in the literature and assume it is independent of the wavelength within the range studied. Figure 4(b) shows the refractive index ($n_b$, $k_b$) of a ~100 nm thick CoFeB film (Si/SiOx/0.5 Ta/100 CoFeB). Here the subscript "b" stands for *bulk* like CoFeB. The refractive index increases with increasing wavelength, typical of metallic films[40]. As the refractive index of a magnetic material typically depends on the polarization of light we compute the refractive index ($n_{\text{CoFeB}}^{\pm}$, $k_{\text{CoFeB}}^{\pm}$) for both polarities ($\sigma^{\pm}$) using the refractive index ($n_b$, $k_b$) and the Kerr rotation angle ($\theta_b$) and ellipticity ($\eta_b$) of the thick CoFeB film. $\sigma^+$ and $\sigma^-$ correspond to right and left circularly polarized lights, respectively. $\theta_b$ and $\eta_b$ are plotted in Fig. 4(c). It is interesting to note that some of the peak values of $\theta_K$ and $\eta_K$ obtained from ~1 nm thick CoFeB layer shown in Figs. 3(a-g) are larger than the corresponding bulk limit shown in Fig. 4(c). $n_{\text{CoFeB}}^{\pm}$ and $k_{\text{CoFeB}}^{\pm}$ are expressed as:

$$n_{\text{CoFeB}}^{\pm} = n_b \pm \Delta n, \; k_{\text{CoFeB}}^{\pm} = k_b \pm \Delta k \qquad (3)$$

$$\Delta n = \left( k_b \operatorname{Re}[\varepsilon_{XY}] - n_b \operatorname{Im}[\varepsilon_{XY}] \right) \big/ \left( n_b^2 + k_b^2 \right)$$

$$\Delta k = \left( n_b \operatorname{Re}[\varepsilon_{XY}] + k_b \operatorname{Im}[\varepsilon_{XY}] \right) \big/ \left( n_b^2 + k_b^2 \right)$$

$$\varepsilon_{XY} = \left[ n_b \left(1 - n_b^2 + 3k_b^2\right) \theta_b - k_b \left(1 - 3n_b^2 + k_b^2\right) \eta_b \right] + i \left[ k_b \left(1 - 3n_b^2 + k_b^2\right) \theta_b + n_b \left(1 - n_b^2 + 3k_b^2\right) \eta_b \right]$$

The calculated refractive index ($n_{\text{CoFeB}}^{\pm}$, $k_{\text{CoFeB}}^{\pm}$) is shown in Fig. 4(d). Although the difference of the refractive index for both polarities is small, such small difference causes the magneto-optical Kerr effect.

We first compute the effective refractive index of the Si substrate and SiOx layer, i.e. ($n_{\text{Si/SiOx}}$, $k_{\text{Si/SiOx}}$). Substituting $n_{\text{SiOx}}$, $k_{\text{SiOx}}$, $n_{\text{Si}}$, $k_{\text{Si}}$ into $n_\alpha$, $k_\alpha$, $n_\beta$, $k_\beta$ of Eq. (1), respectively, the effective refractive index of Si/SiOx ($n_{\text{Si/SiOx}}$, $k_{\text{Si/SiOx}}$) is obtained. Next we compute the effective refractive index of Si/SiOx and the thin film heterostructure (i.e. CoFeB layer). We substitute $n_{\text{CoFeB}}^{\pm}$, $k_{\text{CoFeB}}^{\pm}$, $n_{\text{Si/SiOx}}$, $k_{\text{Si/SiOx}}$ into $n_\alpha$, $k_\alpha$, $n_\beta$, $k_\beta$ of Eq. (1) to obtain the effective refractive index ($n^{\pm}_{Si/SiOx/CoFeB}$, $k^{\pm}_{Si/SiOx/CoFeB}$) of the entire structure. The polarization dependent reflection amplitude of the sample is defined as:

$$r^{\pm} = \frac{(n_0 - ik_0) - (n^{\pm}_{Si/SiOx/CoFeB} - ik^{\pm}_{Si/SiOx/CoFeB})}{(n_0 - ik_0) + (n^{\pm}_{Si/SiOx/CoFeB} - ik^{\pm}_{Si/SiOx/CoFeB})} \quad (4)$$

where $n_0$ and $k_0$ are the real and imaginary parts of the refractive index of air. Here we assume $n_0 \sim 1$ and $k_0 \sim 0$. The calculated average reflection amplitude ($\frac{r^+ + r^-}{2}$) is plotted as a function of light wavelength for different SiOx layer thicknesses in Fig. 2(h-n). The calculated curves resemble those of the corresponding experimental results. To gain insight into the underlying physics of the model calculations, Eq. (4) is rewritten more explicitly as the following:

$$r^{\pm} = \frac{r^{\pm}_{SiOx/CoFeB} + r^{\pm}_{CoFeB/Air} + r_{Si/SiOx}(1 + r^{\pm}_{SiOx/CoFeB} r^{\pm}_{CoFeB/Air})\exp\left[4\pi i(n_{SiOx} + ik_{SiOx})\frac{t_{SiOx}}{\lambda}\right]}{1 + r^{\pm}_{SiOx/CoFeB} r^{\pm}_{CoFeB/Air} + r_{Si/SiOx}(r^{\pm}_{SiOx/CoFeB} + r^{\pm}_{CoFeB/Air})\exp\left[4\pi i(n_{SiOx} + ik_{SiOx})\frac{t_{SiOx}}{\lambda}\right]} \quad (5)$$

We made the following assumption to derive Eq. (5) from Eq. (4), i.e. $\frac{t_{CoFeB}}{\lambda} \ll 1$: this is valid since the thickness of the heterostructure is much smaller than the wavelength. Equation (5) indicates that the reflection amplitude, and consequently the Kerr rotation angle and the ellipticity, will oscillate with the inverse of the light wavelength with a periodicity defined by $\frac{1}{2n_{SiOx}t_{SiOx}} \sim \frac{1}{3t_{SiOx}}$. In Fig. 5(a), we replot the calculated reflection amplitude (Fig. 2(m)) as a function of the inverse of the light wavelength ($\lambda$) for $t_{SiOx}$=500 nm. The blue solid line is a fit to the results using a sinusoidal function with a linear background. The oscillation period of the reflection amplitude agrees with $\sim 1/3t_{SiOx}$.

The Kerr rotation angle and the ellipticity are defined using the reflection amplitude as the following:

$$\theta_K = \frac{1}{2}\left[\arg(r^+) - \arg(r^-)\right]$$
$$\eta_K = \arctan\left[\frac{|r^+|-|r^-|}{|r^+|+|r^-|}\right] \quad (6)$$

The calculated $\theta_K$ and $\eta_K$ spectra for samples with various SiOx layer thicknesses are plotted in Figs. 3(h-n). We find good agreement between the calculation and the experimental results. The wavelength at which peaks in the $\theta_K$ and $\eta_K$ occur coincides with that when the reflection amplitude takes a minimum. To shows this more explicitly, we replot the results of Fig. 3(m) as a function of $1/\lambda$ in Fig. 5(b). The correspondence between the reflection amplitude and the Kerr signals are clear. These results suggest than when the light is confined within the SiOx layer and undergoes multiple reflection, the Kerr signal increases, however, with a reduction in the reflection amplitude.

**Discussion**

Finally we discuss the appropriate SiOx layer thickness for a given light wavelength to obtain the maximum Kerr signal output. We calculate $\theta_K$ as a function of $t_{SiOx}$ to compare with the experimental results. The blue solid line in Fig. 1(b) shows the calculated results. Good agreement between the calculations and experiments are obtained except for $t_{SiOx} \sim 400$ nm (the deviation is also evident in Figs. 3(e) and 3(l) around $\lambda \sim 500$ nm). From the calculations shown in Fig. 1(b), one finds that there are optimum thicknesses for the SiOx layer under a given light wavelength. As the maximum Kerr signal occurs when the reflection amplitude takes a minimum, Eq. (5) indicates that the optimum condition to observe large Kerr signals is $4\pi n_{SiOx} \frac{t_{SiOx}}{\lambda} \sim \pi(2m-1)$, where *m* is an integer. This in turn gives the optimum SiOx thickness $t^*_{SiOx} \sim \frac{\lambda}{n_{SiOx}} \frac{2m-1}{4}$ for a given wavelength. For a green light ($\lambda$~500 nm), $t^*_{SiOx}$ is ~80 nm, 250 nm, 420 nm, 580 nm, which are in good agreement with the SiOx thicknesses at which |$\theta_K$| takes a maximum in the calculations (see Fig. 1(b)). Such interference condition is also required to visualize graphene sheets placed on Si/SiOx substrates using optical microscopes[41].

It should be noted that the enhancement of $\theta_K$ is associated with a reduction in the reflected light amplitude. This is because multiple reflections occur at the Si/SiOx and SiOx/heterostructure interfaces under the interference condition, and as a result, there is a larger probability of the light being absorbed into the Si substrate and/or the heterostructure, resulting in a reduction of the light amplitude that reflects off the sample. The correspondence can be seen both in the experimental results and the simulated curves (see Figs. 2 and 3). Significant effort was placed in the past to maximize the Kerr rotation angle while maintaining a reasonable amount of reflected light amplitude to detect the magnetic state of the magneto-optic media[35].

We note that the optical interference effect may provide a useful means to study the interface states with magneto-optics. A common approach taken thus far is to use magnetic multilayers[42-44] in which a large number of interfaces contribute to the overall signal and thereby increase the signal to noise ratio. Instead, one may use the optical interference within the oxide layer of the substrate and probe, for example, a single interface within the film many times via the multiple reflections associated with the interference. As we have shown here, even for an ultrathin film almost transparent to visible light (~1.5 nm thick metallic film), one can use conventional models[13,36] to calculate the expected magneto-optical signal. Thus, this will allow a straightforward way to evaluate other contributions, if any, to the signal that may result from e.g. interface spin orbit coupling.

Experimentally, we find that the measured Kerr signals (rotation angle and ellipticity) at the resonance condition (when $t_{SiOx} \sim t_{SiOx}^*$) are in some cases larger in magnitude that those from the calculations. For example, the measured Kerr ellipticity at $t_{SiOx} \sim 100$ nm and $\lambda \sim 630$ nm (Fig. 3(b)) is nearly twice as large as that of the calculations (Fig. 3(i)). The reflection amplitude, in contrast, is better reproduced by the calculations. These results suggest that the Kerr signal of the heterostructure with the ultrathin magnetic film is enhanced compared to what is expected from the corresponding bulk properties. We infer that spin orbit coupling, magnetic moment, and/or magneto-optical coupling at the interface(s)[44] may be enhanced in the heterostructure, leading to larger Kerr signals.

In summary, we have studied conditions at which one can maximize the magneto-optical Kerr effect from an ultrathin magnetic layer, almost transparent to visible light, deposited on Si substrate coated with a SiOx layer. For a given light wavelength ($\lambda$), the thickness of the SiOx layer has to be set to one quarter of the effective wavelength ($\lambda/n_{SiOx}$) times an odd integer to observe the maximum Kerr signal ($n_{SiOx}$ is the real part of the SiOx refractive index). Such condition fulfills the requirement to cause optical interference within the SiOx layer. Interestingly, we find that the Kerr signals are, for certain resonance conditions, larger than the estimation despite the good agreement of the measured and calculated reflection amplitude. The discrepancy may originate from interface states that are distinct from the bulk characteristics. These results show that optical interference effect can be utilized to study the magneto-optical properties of ultrathin films, including those of the interface states.

**Methods**

*Sample preparation*

RF magnetron sputtering is used for the film deposition. The substrate is non-doped Si(001) substrates coated with SiOx. Wet oxidation is used for the formation of SiOx. The SiOx thickness is determined by ellipsometry and its accuracy is ~±10%. For the growth of heterostructures, relatively thick (~20 nm) calibration films are made to estimate the film deposition rate. Line profiling and x-ray resonant reflection are used to measure the thickness of the calibration films. The variation of the measured deposition rate is typically less than ~10%.

*Ellipsometry*

Ellipsometry from JASCO (model M-150) is used. The angle of incidence is ~45 degree. A linearly polarized light is fed into a photo-elastic modulator (PEM) and the output from the PEM is irradiated on the sample. No magnetic field is applied during the ellipsometry measurements. The magnetization direction of the ~100 nm thick CoFeB film points along the film plane.

*Kerr spectra*

Magneto-optical Kerr spectrum is measured using a Xenon light source, a PEM and a photomultiplier detector. The angle of incidence is ~4 deg. Details of the setup are described in ref. [45].

**Acknowledgements**

This work was partly supported by JSPS Grant-in-Aids for Scientific Research (16H03853), Specially Promoted Research (15H05702), MEXT Nanofab platform, the MEXT-Supported Program Research Foundation at Private University (2014-2020), and Spintronics Research Network of Japan.

**Author contributions**

M.H. and H.A. planned the study. M.H. prepared the films, S.S. measured the Kerr spectra and analyzed the results with inputs from H.A. and M.H. All authors have discussed the results and commented on the manuscript.

**Additional Information**

**Competing financial interests**: The authors declare no competing financial interests.

**Figure captions**

**Fig. 1**. (a) Kerr signal ($A$) vs. out of plane magnetic field ($H_Z$) for a film stack (Sub./0.5 Ta/1 CoFeB/2 MgO/1 Ta) deposited on Si substrates with different SiOx thicknesses. The curves are shifted vertically by 1 deg for clarity. (b) The image shows a schematic illustration of the sample. The solid circles in the plot shows the Kerr rotation angle ($\theta_K$) as a function of the SiOx thickness. The laser wavelength ($\lambda$) is ~500 nm. The solid line represents calculation results using Eqs. (5) and (6).

**Fig. 2**. (a-n) Experimentally measured (a-g) and analytically calculated (h-n) reflection amplitude vs. laser wavelength for a film stack (Sub./0.5 Ta/1 CoFeB/2 MgO/1 Ta) deposited on Si substrates with different SiOx thicknesses.

**Fig. 3**. (a-n) Experimentally measured (a-g) and analytically calculated (h-n) Kerr rotation angle $\theta_K$ (black squares) and Kerr ellipticity $\eta_K$ (red circles) vs. laser wavelength for a film stack (Sub./0.5 Ta/1 CoFeB/2 MgO/1 Ta) deposited on Si substrates with different SiOx thicknesses.

**Fig. 4**. (a) Light wavelength dependence of the real ($n_{Si}$) and imaginary ($k_{Si}$) parts of the refractive index for Si substrate without SiOx. (b,c) Real ($n_b$) and imaginary ($k_b$) parts of the refractive index (b), Kerr rotation angle $\theta_b$ and ellipticity $\eta_b$ (c) of a ~100 nm thick CoFeB film (Sub./0.5 Ta/100 CoFeB) deposited on a Si substrate with no SiOx. (d) Calculated real ($n_{CoFeB}^{\pm}$)

and imaginary ($k_{CoFeB}^{\pm}$) parts of the refractive index of the thick CoFeB film for different light polarization. Eq. (3) is used to obtain these results.

**Fig. 5**. (a,b) Calculated reflection amplitude (a), Kerr rotation angle and ellipticity (b) plotted against the inverse of the light wavelength $\lambda$. The blue line in (a) shows fit to the data using a sinusoidal function. The thickness of the SiOx layer is 500 nm.

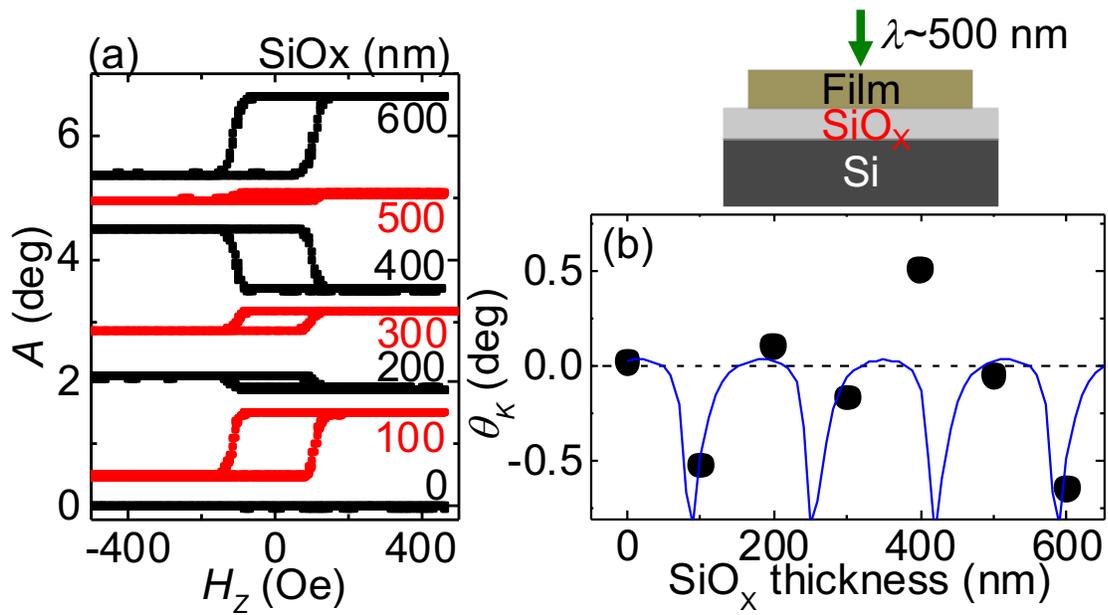

Fig. 1

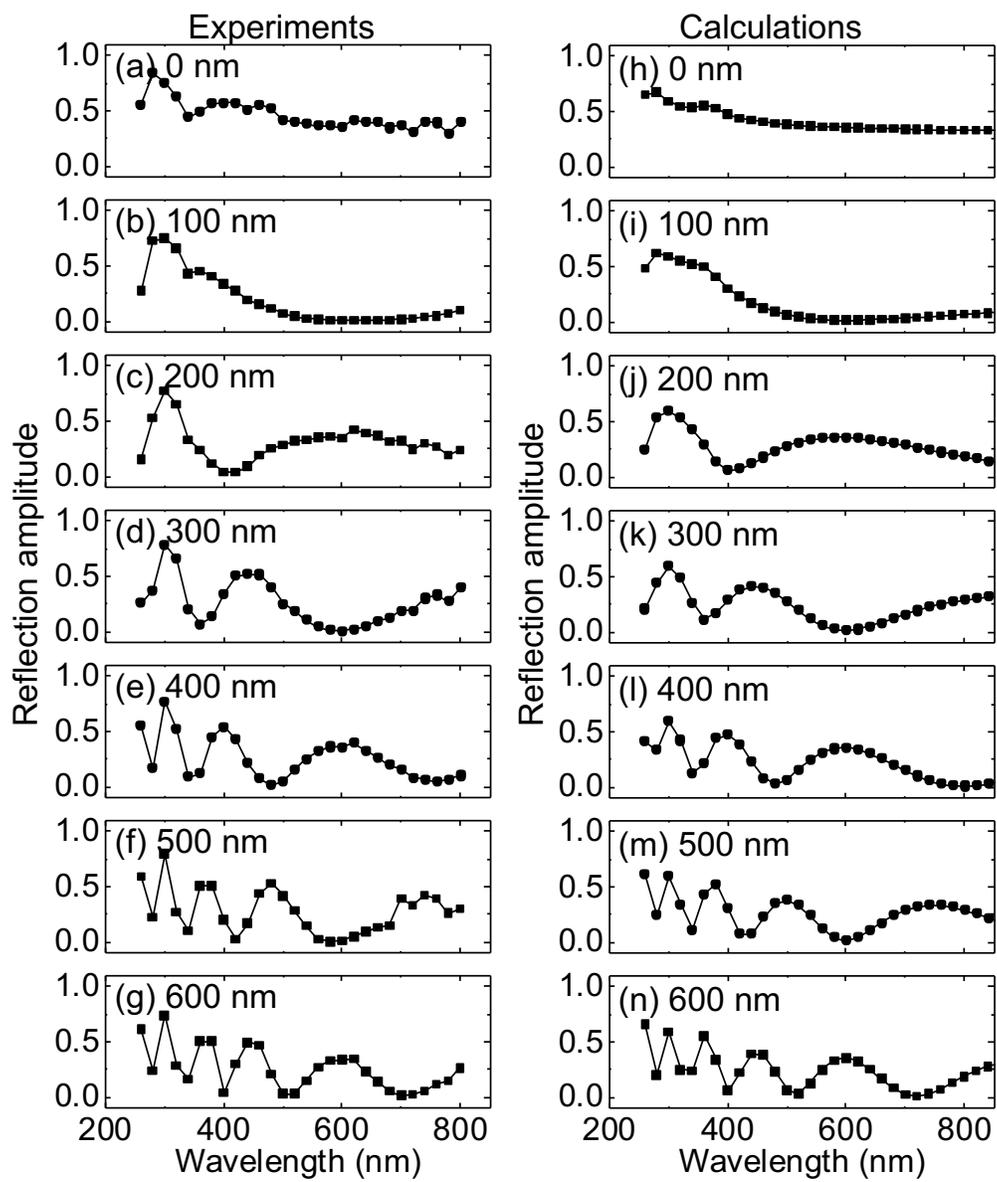

Fig. 2

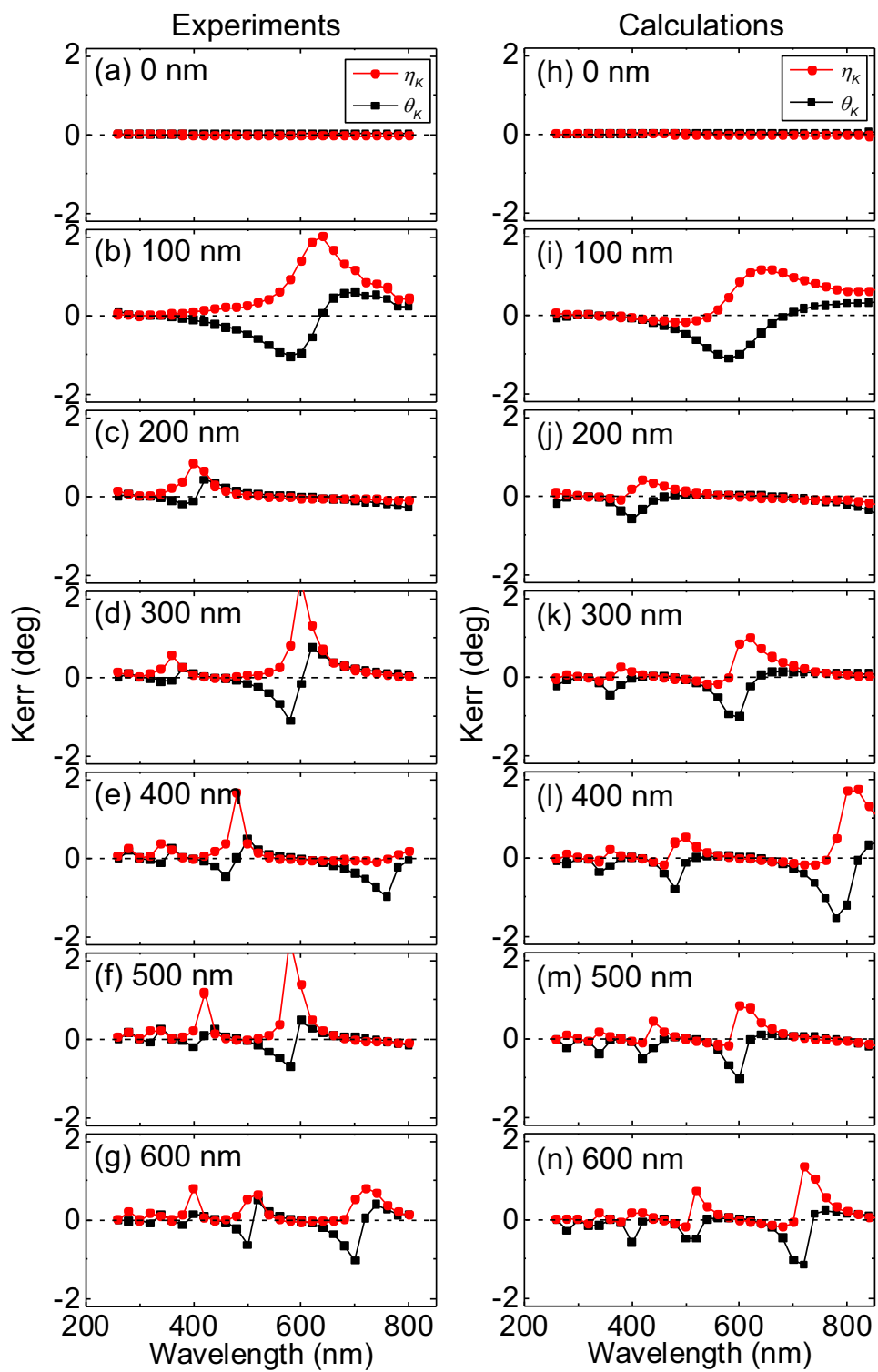

Fig. 3

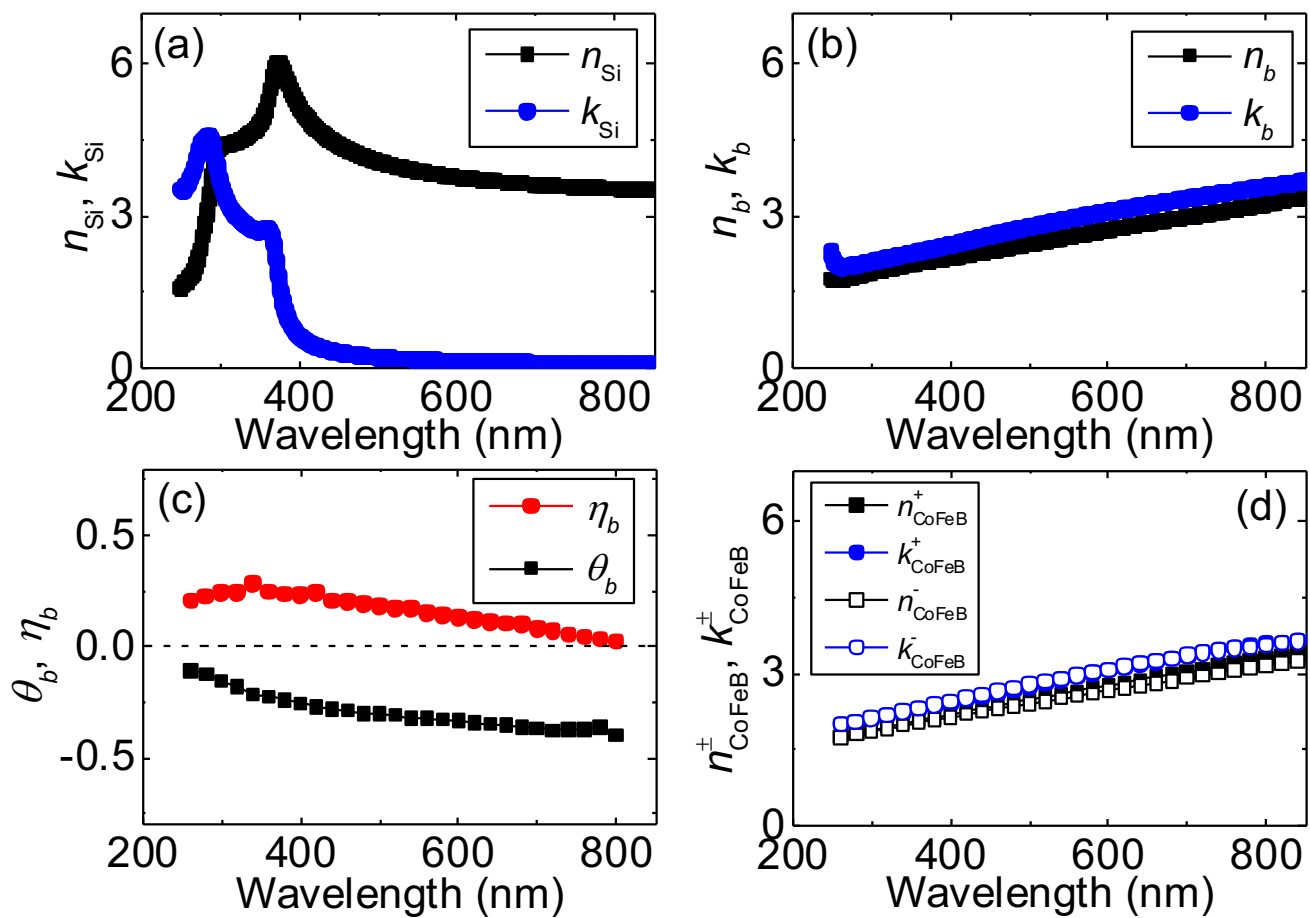

Fig. 4

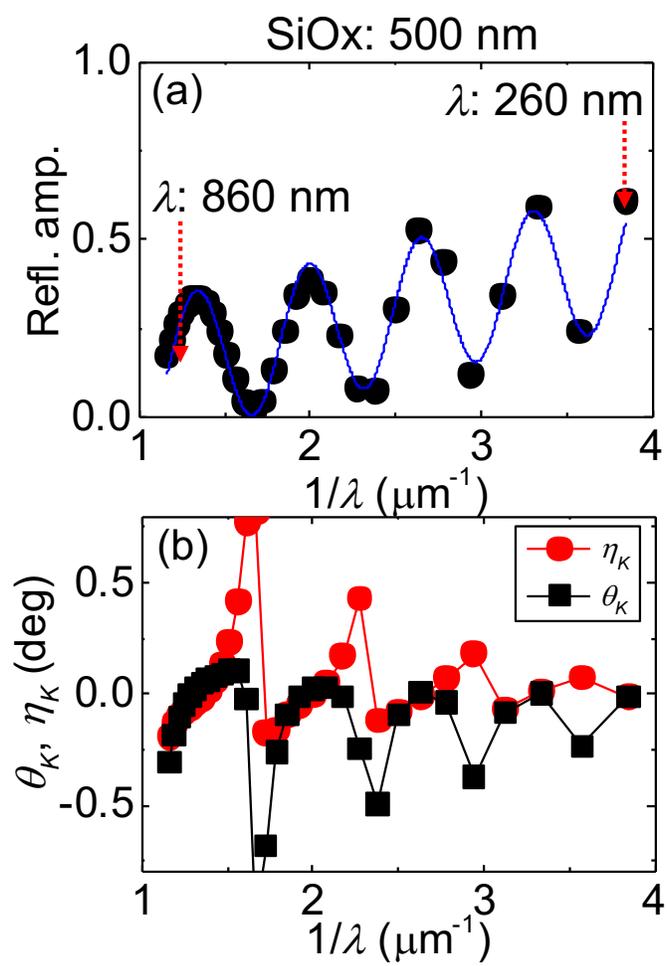

Fig. 5